%% file: main.tex
\documentclass[10pt,twocolumn]{article}
\usepackage[twoside=true,head=13pt,paperwidth=8.5in,paperheight=11in,columnsep=2pc,top=57pt,bottom=73pt,inner=54pt,outer=54pt,marginparwidth=2pc,heightrounded]{geometry}

\usepackage[colorinlistoftodos]{todonotes} %
\usepackage{subcaption} %
\usepackage{amsmath} %
\usepackage{longtable} %
\usepackage{colortbl} %
\usepackage{hyphenat} %
\usepackage{fancyvrb} %
\usepackage{adjustbox} %
\usepackage{xurl} %
\usepackage[hidelinks]{hyperref}
\title{\textbf{Implementing Data Diodes Using Commodity Hardware and Open Source Software}}

\author{
  \begin{minipage}[t]{0.45\textwidth}
    \centering
    Peter Story\\
    Clark University\\
    Worcester, MA, USA\\
    \texttt{PeStory@clarku.edu}
  \end{minipage}
  \hfill
  \begin{minipage}[t]{0.45\textwidth}
    \centering
    Gert-Jan den Besten\\
    Dutch Ministry of Defence\\
    Netherlands\\
    \texttt{gertjan.den.besten@gmail.com}
  \end{minipage}
}
\date{}

\begin{document}

\twocolumn[
  \begin{@twocolumnfalse}
    \maketitle
  \end{@twocolumnfalse}
]

\begin{abstract}
One-way network devices, known as data diodes, are used to defend against sophisticated cyberattacks.
Partly due to their high cost, data diodes are mostly deployed in nuclear power plants and within the government for handling classified information.
Although commercially available data diodes are expensive, a data diode's hardware can assembled from commodity fiber-optic network equipment.
However, specialized software is needed to send data through a data diode reliably: the receiving program cannot request retransmission of dropped packets, so packet loss must be minimized and mitigated.
First, we developed a minimal program to measure packet loss.
We found that most packet loss was caused by the receiving program processing incoming packets too slowly, and that packet loss often occurs in clusters.
Also, we discovered ways to minimize packet loss on Linux and macOS without using superuser privileges.
Next, we tested three existing open source programs for one-way data transfers: netcat, UDPcast, and lidi.
Although these programs were unreliable in their default configurations, we identified reliable configurations for UDPcast and lidi.
Finally, we incorporated our findings into pydiode, our cross-platform program for reliable one-way data transfers.
\end{abstract}

\maketitle

\input{1-introduction}
\input{2-related-work}
\input{4-method}
\input{5-results}
\input{6-limitations}
\input{7-discussion}
\input{8-conclusions}

\bibliographystyle{plain}
\bibliography{manual,zotero}

\clearpage
\onecolumn

\appendix

\input{9-appendix}

\end{document}

%% file: 1-introduction.tex
\section{Introduction}

It is challenging to defend against technically sophisticated cyberattacks.
Militaries, intelligence agencies, and other powerful entities use unpatched software vulnerabilities to launch \textit{zero-day attacks} against high-value targets.
Nearly any internet-connected device can be hacked by zero-day attacks~\cite{andersonWhyInformationSecurity2001,fidlerZeroProgressZero2024}.
Software-based solutions offer limited protection, since they can be hacked themselves.
One approach to defend against zero-day attacks is to physically isolate devices from the internet using an \textit{air gap}.
However, air gaps are often impractical, since it is necessary to exchange data with most devices.
Manually transferring data across an air gap using a USB drive is cumbersome and insecure~\cite{berretGuideSecureDrop2016,luBADUSBCRevisitingBadUSB2021,voutevaFeasibilityDeploymentBad2015,blanchetBadUSBThreatHidden2018}.
One-way network devices, known as \textit{data diodes}, offer a more secure and usable alternative~\cite{cohenDesigningProvablyCorrect1988,stevensImplicationOpticalData1999}.
Data diodes are physically limited to only transfer data in one direction.
Today, data diodes are mostly deployed in nuclear power plants and within the government for handling classified information.
Commercially available data diodes are expensive, with some costs exceeding one hundred thousand dollars~\cite{barryDataDiodesCyber2012}.
Their high prices may discourage more widespread adoption.
A data diode's hardware can be assembled from commodity fiber-optic network equipment~\cite{arnoldStrategiesTransportingData2016,OSDD,storyBuildingAffordableData2023}.
However, specialized software is needed to send data through a data diode reliably: the receiving program cannot request retransmission of dropped packets, so packet loss must be minimized and mitigated.
We answer three research questions about data diodes:
\begin{enumerate}
\item What causes packet loss in data diodes?
\item How does existing open source software for one-way data transfers perform? In particular, how long do transfers take, and how often do transfers fail?
\item Can our own software send data reliably in less time than existing software?
\end{enumerate}

First, we developed a minimal program to measure packet loss under different conditions on Linux and macOS (\S~\ref{sec:sw_dev}).
We collected statistics on packet loss while transferring terabytes of data through a data diode built using commodity network hardware (\S~\ref{sec:minimal_linux_results} and \S~\ref{sec:minimal_macos_results}).
In our testing, 99.99\% of packet loss was due to RcvbufErrors, caused by the receiving program processing incoming packets too slowly.
We discovered several ways to minimize packet loss without using superuser privileges.
Some results were expected: packet loss can be minimized by using a separate helper thread to write out received data, programs on Linux can send packets with larger payloads, and programs on macOS can request a larger receive buffer size.
However, we also made several counterintuitive discoveries.
First, on both Linux and macOS, we observed cases where transferring data more slowly \textit{decreased} transfer reliability.
Second, we found that although sending larger packets increased reliability on Linux, sending larger packets on macOS could \textit{decrease} reliability.

Next, we performed one-way data transfers using three existing open source programs: netcat, UDPcast, and lidi.
We found that only 93.89\% of netcat's transfers succeeded (\S~\ref{sec:netcat_results}).
Transfers were also unreliable when using UDPcast (\S~\ref{sec:udpcast_results}) and lidi (\S~\ref{sec:lidi_results}) in their default configurations.
However, we discovered configurations for UDPcast and lidi that enabled reliable transfers using these programs.
We also incorporated our findings into the development of pydiode, a cross-platform program for reliably transferring information through data diodes (\S~\ref{sec:sw_dev}).
pydiode uses redundancy to mitigate packet loss and can automatically detect transfer errors.
Unlike UDPcast and lidi, our minimal program and pydiode do not employ forward error correction (FEC).
Nevertheless, both minimal and pydiode offer faster reliable transfers than UDPcast, and they are only 52\% and 46\% slower than lidi, respectively (Table~\ref{table:summary}).
When properly configured, the choice between pydiode, UDPcast, and lidi may depend on factors beyond transfer speed (\S~\ref{sec:comparison}).
For example, lidi's FEC implementation may infringe on RaptorQ patents held by Qualcomm, whereas pydiode and UDPcast's underlying technology is public domain.
Finally, we offer advice for software developers (\S~\ref{sec:advice}), and describe opportunities for future work (\S~\ref{sec:conclusions}).

%% file: 2-related-work.tex
\section{Related Work}
\label{sec:related_work}
First, we describe the threat posed by zero-day attacks, which are used by powerful entities against high-value targets (\S~\ref{sec:zero_day}).
Next, we explain how air gaps and one-way communication (i.e., data diodes) can defend against zero-day attacks (\S~\ref{sec:air_gap_data_diode}).
Currently, data diodes are mainly used by the military and in nuclear power plants, in part due to their high cost (\S~\ref{sec:existing_deployments}).
Finally, we survey existing open source software for data diodes (\S~\ref{sec:open_source_software}).

\subsection{Zero-Day Attacks}
\label{sec:zero_day}

Cybersecurity threats vary in technical sophistication.
The most advanced attacks can succeed without user interaction, even if the target device has all software updates installed.
Commercial spyware exploits software vulnerabilities to take over targets' devices, and is available to governments through products like the NSO Group's Pegasus~\cite{bergmanBattleWorldMost2022} and Cytrox's Predator~\cite{goodinIOS0daysCellular2023}.
Software-based solutions such as firewalls, sandboxing, code signing, and executable-space protection increase the cost of attacks, but cannot completely prevent attacks; software-based solutions can be evaded by combining vulnerabilities, or can be hacked themselves.

The economics of cybercrime have protected most users from technically sophisticated attacks.
Although software vulnerabilities are widespread in modern software, discovering vulnerabilities takes time and expertise.
If a software vulnerability is widely exploited, software vendors will patch their software to close the vulnerability.
Thus, at a given time there is a limited supply of unpatched vulnerabilities, also known as \textit{zero-days}~\cite{ablonZeroDaysThousands2017,fidlerZeroProgressZero2024}.
Because zero-days are valuable~\cite{zerodiumArchive,fidlerZeroProgressZero2024}, attackers save them for high-value targets.
However, artificial intelligence is accelerating the discovery of zero-day exploits~\cite{anthropicProjectGlasswingSecuring2026,schneierHowAIChanging2026}, potentially exposing more organizations to zero-day attacks.

\subsection{Air Gaps and Data Diodes}
\label{sec:air_gap_data_diode}
Defending against zero-day attacks is incredibly challenging~\cite{andersonWhyInformationSecurity2001,ablonZeroDaysThousands2017}.
One way high-value targets defend against zero-day attacks is by physically isolating devices using an \textit{air gap}.
In theory, if an attacker cannot communicate with a device, they cannot hack it.
In practice, many air-gapped devices are not completely isolated, as there is a need to exchange some data with the device.
However, if the direction of the flow of information is controlled, guarantees can be made about confidentiality~\cite{bellSecureComputerSystems1973,bellLookingBackBellLa2005} and integrity~\cite{kennethIntegrityConsiderationsSecure1977}.
In particular, the Bell–LaPadula model (BLP) implies that if information cannot leave a system containing confidential information, then confidentiality is guaranteed even if that system is compromised~\cite{bellSecureComputerSystems1973,bellLookingBackBellLa2005}.
Similarly, the Biba Integrity Model implies that if a high integrity system cannot receive information from lower integrity systems, then lower integrity systems cannot compromise the high integrity system~\cite{kennethIntegrityConsiderationsSecure1977}.

The direction of information flow can be enforced using devices which are physically limited to only transfer data in one direction, known as \textit{data diodes}~\cite{cohenDesigningProvablyCorrect1988,stevensImplicationOpticalData1999}.
A data diode is a network device which is only physically capable of transferring data in one direction.
A network firewall is not a data diode, since a software vulnerability could compromise the firewall and modify its rules.
In contrast, a data diode should include hardware that only allows data to be transferred in one direction.
For example, using a modified fiber-optic link~\cite{arnoldStrategiesTransportingData2016,OSDD,storyBuildingAffordableData2023}.
These physical properties allow for guarantees about the direction of data flow~\cite{cohenDesigningProvablyCorrect1988,stevensImplicationOpticalData1999}.

\subsection{Existing Deployments of Data Diodes}
\label{sec:existing_deployments}
\begin{figure*}
\centering
\begin{subfigure}[t]{.49\linewidth}
\centering
\includegraphics[width=\linewidth]{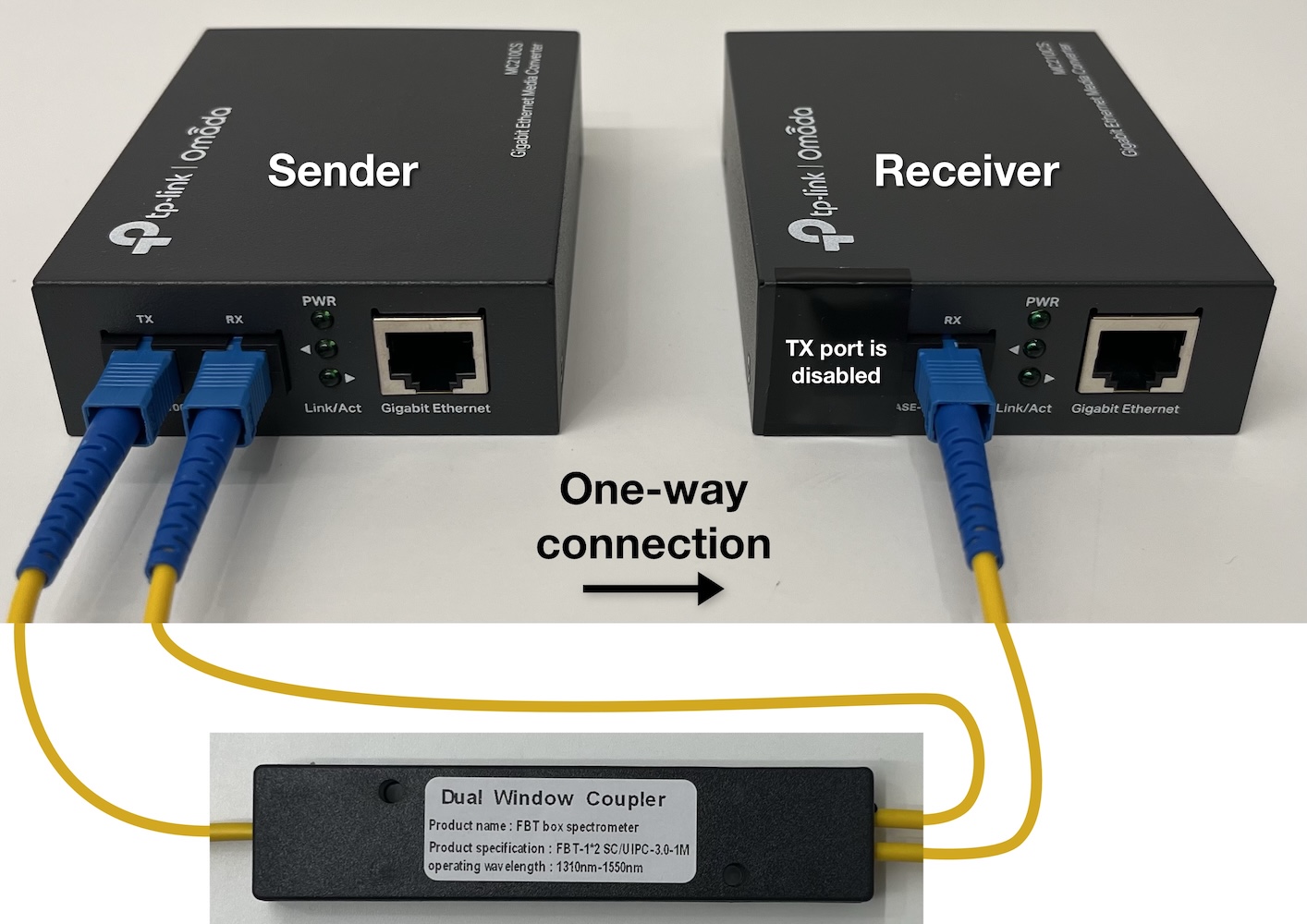}
\caption{Our data diode is built using fiber-optic media converters. The converter on the left sends data to the converter on the right. The converter on the right physically cannot transfer data in the reverse direction, since its transmit port is taped over.}
\label{fig:diode}
\end{subfigure}%
\hfill
\begin{subfigure}[t]{.49\linewidth}
\centering
\includegraphics[width=.9\linewidth]{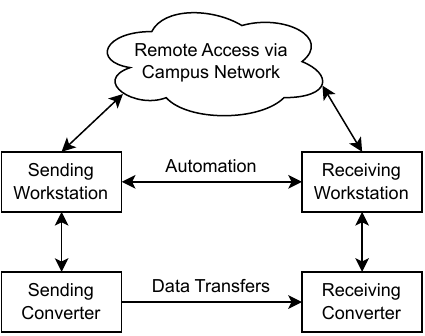}
\caption{Our workstations used separate network connections for remote access, experiment automation, and data transfers. Bidirectional arrows depict standard Ethernet connections. The one-way arrow depicts a one-way connection, enforced by the data diode.}
\label{fig:workstations}
\end{subfigure}
\caption{Network configuration for our experiments}
\label{fig:network}
\end{figure*}

Data diodes are primarily deployed in two different contexts to protect confidentiality and integrity, respectively.
First, data diodes are used by military and intelligence agencies as a \textit{cross-domain solution (CDS)}, to limit exchange of data between different security classification levels~\cite{arnoldStrategiesTransportingData2016}.
For example, a data diode can ensure that data only flows from a lower classification network to a higher classification network.
Second, data diodes are deployed in certain safety-critical public sector industries~\cite{industrialcontrolsystemscyberemergencyresponseteamRecommendedPracticeImproving2016}.
For example, data diodes are mentioned frequently in the U.S. Nuclear Regulatory Commission's regulations for nuclear power plants~\cite{advisorycommitteeonreactorsafeguardsdigitalinstrumentationandcontrolOfficialTranscriptProceedings2021,bergemannCyberSecurityEvent2015,downsCyberSecurityPrograms2017,officeofnuclearregulatoryresearchCyberSecurityPrograms2010}.
Literature from data diode vendors describe deployments in other industries as well~\cite{fendIndustries,owlcyberdefense,siemens}.
However, a recent industry report found that data diodes had lower adoption than other cybersecurity technologies, and suggested this may be due to their ``reputation for complexity and cost''~\cite{harpCS2AIKPMGControlSystem2024}.
Indeed, proprietary data diodes can have prices exceeding one hundred thousand dollars~\cite{fendArchive,larkinSecuringPhotovoltaicSystem2020,barryDataDiodesCyber2012}, and each product offers different features.
Alternatively, a data diode's hardware can be assembled from commodity fiber-optic network equipment for less than \$100~\cite{arnoldStrategiesTransportingData2016,OSDD,storyBuildingAffordableData2023}, as shown in Figure~\ref{fig:diode}.
Still, specialized software is needed to send data through a data diode reliably.

\subsection{Open Source Software for Data Diodes}
\label{sec:open_source_software}

\begin{table*}
\centering
\begin{adjustbox}{max width=1\linewidth}
\begin{tabular}{| p{2.5cm} | p{5cm} | p{4cm} | p{4cm} |}
\hline
\textbf{Program} & \textbf{Functionality to Increase the Reliability of Transfers} & \textbf{Input Format} & \textbf{Platforms} \\
\hline
netcat~\cite{netcat} & No extra functionality to increase the reliability of transfers. & Standard input stream (or file via redirection) & Different implementations for Linux, macOS, and Windows \\ \hline
UDPcast~\cite{udpcast} & Rate limiting and forward error correction based on Vandermonde matrices~\cite{rizzoEffectiveErasureCodes1997}. & Standard input stream or file & Linux, Windows \\ \hline
lidi~\cite{lidi} & Forward error correction based on RaptorQ codes~\cite{minderRaptorQForwardError2011,qualcommRaptorQTechnicalOverview2010}. & Standard input stream, TCP stream, UNIX stream, file & Linux \\ \hline
hairgap~\cite{hairgap} & Rate limiting, redundancy, and forward error correction based on wirehair~\cite{wirehair}. & Standard input stream (or file via redirection) & Linux \\ \hline
godiode~\cite{godiode} & Rate limiting, checksums, and redundancy. & Directory of files & Linux, macOS \\ \hline
BlindFTP~\cite{lagadecDiodeReseauExeFilter2006} & Rate limiting, checksums, and redundancy. & Directory of files & Linux, macOS, Windows \\ \hline
Sven Seeberg's data diode~\cite{seeberg} & Rate limiting, checksums, and redundancy. & UDP packets, directory of files & Linux, OpenBSD \\ \hline
\end{tabular}
\end{adjustbox}
\caption{A summary of command-line programs for transferring data over a data diode using UDP.}
\label{table:software}
\end{table*}

Data diodes ensure that data only flows in one direction, necessitating the use of a connectionless protocol like UDP.
Furthermore, the receiver cannot request retransmission of dropped packets, so packet loss must be minimized and mitigated.
As shown in Table~\ref{table:software}, a variety of open source software is capable of transferring data through data diodes using UDP.
We identified this software by searching on Google and GitHub; although the list is not necessarily exhaustive, it includes the most commonly used programs, along with some lesser-known software.
The software's documentation varies in maturity, so we reviewed the documentation and source code to determine each software's functionality and supported input formats.
When documentation listed supported platforms or when precompiled binaries were available, we list those platforms; otherwise, we list the platforms on which we were able to compile and run the software.

Note that stream-based input formats are the most flexible, since they allow sending individual files via file redirection, directories of files using tar, or network traffic using a proxy program.
We chose to evaluate netcat, UDPcast, and lidi because all support transferring data streams.
We considered testing hairgap, but we could not compile it, and the project is not actively maintained.

%% file: 4-method.tex
\section{Method}
\label{sec:method}
Data diodes include hardware for enforcing one-way network traffic, and software for transferring data reliably.
First, we describe how we assembled a data diode, and how we connected two workstations for testing (\S~\ref{sec:hardware}).
Next, we explain our methodology for testing netcat, UDPcast, and lidi (\S~\ref{sec:sw_config}).
Finally, we describe our implementation of a minimal test program (``minimal'') and a more full-featured program (``pydiode'') (\S~\ref{sec:sw_dev}).

\subsection{Hardware Configuration}
\label{sec:hardware}

\subsubsection*{Data Diode Assembly}
A data diode can be built using fiber-optic network equipment.
We assembled our data diode according to the OSDD project's instructions~\cite{OSDD}.
We used two Gigabit Ethernet copper to single mode fiber media converters (TP-Link MC210CS) connected by an SC/UPC single mode fiber-optic splitter.
As of spring 2025, the components cost about \$70 on Amazon.
Figure~\ref{fig:diode} illustrates how the components are connected.
The splitter's input is connected to the TX port of the sender.
One of the splitter's two outputs should be connected to the sender's RX port, and the other to the receiver's RX port.
Without the loop from the sender's TX to its RX, Ethernet autonegotiation will fail, which will prevent data transfer~\cite{autonegotiation,sendOnlyEthernet}.
Finally, we covered the TX port of the receiver with electrical tape.

\subsubsection*{Workstation Configuration}
For Linux testing, we used two Intel NUC 12 mini PCs (NUC12WSHi5).
Each PC had 64~GB of DDR4 RAM and 1~TB of PCIe storage.
Figure~\ref{fig:workstations} shows our PCs' network configurations.
We connected each PC to one of the media converters, designating one PC the sender and the other the receiver.
To automate data collection, we also connected the PCs directly using an Ethernet cable, allowing the sender to control the receiver over SSH without traversing our campus network.
For remote management, both PCs were also connected to our campus network using an additional wired connection.
Our software configuration ensured that data transfers used the data diode, which we confirmed by monitoring the activity lights on the media converters.

For macOS testing, we used two M1 Mac minis.
Each Mac mini had 16~GB of RAM and 512~GB of storage.
We connected the Mac minis the same way we connected the PCs (Figure~\ref{fig:workstations}).

\subsection{Software Configuration}
\label{sec:sw_config}
We performed tests on Linux and macOS.
Linux can be used in industrial deployments of data diodes.
Data diodes can also be used to protect mobile devices against malware~\cite{storyDefendingMessagingApps2026}.
Mobile devices overwhelmingly run Android and iOS, which share network stacks with Linux and macOS, respectively.
Thus, our findings can inform future work testing with mobile devices.

For Linux testing, we installed Ubuntu Desktop 24.04.3 LTS on both PCs.
After installing all available updates in December 2025, we disabled automatic updates.
Python 3.12.3 was installed by default, so we used this version to run our Python code.
Both PCs ran Linux kernel 6.14.0-35-generic.
We moved the interfaces connected to the data diode into network namespaces and assigned IP addresses in the 10.0.1.0/24 range.
Using network namespaces ensured that each data transfer program used the data diode.
Network namespaces also prevented other programs on the system from sending traffic through the data diode, allowing us to collect accurate counts of packets sent and received.
We also added a manual ARP entry (Address Resolution Protocol) for the receiver's IP address, since ARP cannot resolve without bidirectional communication.

We installed macOS Sequoia 15.7.4 on both Mac minis.
After installing all available updates in March 2026, we disabled automatic updates.
Python 3.9.6 was installed by default, but this version of Python does not include all the standard library features we needed.
For consistency with Ubuntu, we used MacPorts to install Python 3.12.12, which we used to run our Python code.
macOS does not support network namespaces, but similar to Linux we assigned IP addresses in the 10.0.1.0/24 range to the interfaces connected to the data diode.
Also, we increased the maximum UDP packet size to allow sending packets of the same size as on Linux, and we added a manual ARP entry for the receiver's IP address.
Finally, we excluded our experiment logs from Spotlight search indexing and we disabled video wallpapers.

Prior work suggests that increasing kernel network buffer sizes and process priority can increase the reliability of transfers~\cite{stevensImplicationOpticalData1999,pietre-cambacedesDeconstructionIndustrialControl2009,linResearchPacketLoss2013}.
However, these changes can require superuser privileges, making them incompatible with certain contexts.
Linux and macOS default to 213~kB and 787~kB UDP receive buffers, respectively.
On Linux, UDP buffers cannot be increased beyond this default without superuser privileges, whereas programs on macOS can request buffers up to 8.4~MB.
Increasing the kernel's network buffer size beyond these limits requires superuser privileges on Linux and macOS, and is impossible on Android and iOS.
Process priority can be configured directly using superuser privileges on Linux and macOS.
Without superuser privileges, process priority can only be adjusted within limits on Android~\cite{androidThreadPriority}, iOS, and macOS~\cite{DispatchQoS}.
To increase the generalizability of our findings, we focused on the configuration options available without superuser privileges.
We did use superuser privileges for three purposes.
First, we used superuser privileges to use network namespaces on Linux.
To confirm that using network namespaces did not affect the reliability of transfers, we reran some experiments without them, and we verified that the results were substantially the same.
Second, we used superuser privileges to add a manual ARP entry on Linux and macOS.
UDPcast and lidi fail to send data without a manual ARP entry, whereas netcat and our own programs support sending to a broadcast address (e.g., 10.0.1.255).
However, in principle it should be possible to modify UDPcast and lidi to support sending to broadcast addresses.
Third, when testing minimal on Linux and macOS, we used superuser privileges to increase the receive buffer size and maximum UDP payload size, respectively.
On Linux, we increased the receive buffer size to 8.4~MB, for comparison to macOS.
On macOS, we increased the maximum UDP payload size to 65,507 bytes, for comparison to Linux.
By default, macOS only allows maximum payload sizes of 1,472 and 9,216 bytes for broadcast and unicast packets, respectively.\footnote{UDP packets with payloads exceeding 1,472 bytes will be fragmented across multiple standard Ethernet frames, which have a 1,500 byte maximum transmission unit (MTU). Network hardware often supports Ethernet jumbo frame MTUs up to 9,216 bytes. IPv4 limits UDP payloads to 65,507 bytes.}
Our results show how to achieve reliable transfers even without superuser privileges.

Our goal was to determine the optimal configuration for each data transfer program.
Specifically, the configuration with minimum data transfer duration for which all transfers would succeed.
We tested the software by repeatedly transferring data through the data diode, and comparing the checksums of the sent and received data.
We chose to transfer 1~Gbit ($10^9$~bits) of randomly generated data in each trial.
This value was large enough to exceed the default network buffer size, and small enough to allow testing many transfers in a reasonable amount of time.
Figure~\ref{fig:loop} in Appendix~\ref{sec:figures} describes our experimental design in pseudocode.
As part of each transfer, we also recorded network statistics (e.g., the number of packets sent and received), which we used to diagnose the reasons for packet loss.
We power cycled the workstations between each experiment.
Next, we explain the configurations we tested for netcat, UDPcast, and lidi.

\subsubsection*{Configuring netcat}
We tested the OpenBSD rewrite of netcat, available in the netcat-openbsd package, since it is installed by default in Ubuntu.
netcat does not offer any options to increase the reliability of transfers.
Figure~\ref{fig:netcat_cmds} in Appendix~\ref{sec:figures} shows the commands we used to test netcat.

\subsubsection*{Configuring UDPcast}
We tested UDPcast version 20120424-2build1.
Although newer releases are available on UDPcast's website~\cite{udpcast}, we tested the latest version available from Ubuntu's package manager.
Figure~\ref{fig:udpcast_cmds} in Appendix~\ref{sec:figures} shows the commands we used to test UDPcast.
In our testing, we varied the transfer rate and forward error correction (FEC) settings.
FEC is implemented using an erasure code based on Vandermonde matrices~\cite{rizzoEffectiveErasureCodes1997}.
The FEC setting specifies how many ``stripes'' each chunk of data is split into, the number of FEC packets included with each stripe, and the number of data packets in each stripe~\cite{udpcastCMD}.
For example, with \texttt{-{}-fec 8x16/128}, each chunk of data will be sent in 8 stripes, and each stripe will have 16 FEC packets and 128 data packets.

\subsubsection*{Configuring lidi}
We tested lidi v2.1.0.
Figure~\ref{fig:lidi_cmds} in Appendix~\ref{sec:figures} shows the commands we used to test lidi.
Although lidi's transfer rate is not configurable, we varied lidi's FEC settings.
lidi implements FEC using RaptorQ codes~\cite{minderRaptorQForwardError2011,qualcommRaptorQTechnicalOverview2010}.
lidi's \texttt{-{}-repair} argument controls the amount of repair data, specified as a percent of the original data~\cite{lidiCLI}.
By default, lidi sends repair data equal to 2\% of the original data.
Error correction data is associated with a block of original data, with a default size of 734,928 bytes.
We observed that when lidi's transfers fail, the receiving program never exits.
Thus, we used to the \texttt{timeout} command to terminate the receiving program after 60 seconds.
When analyzing the results, we counted these transfers as failures with undefined durations.

\subsection{Software Development}
\label{sec:sw_dev}
First, we developed a minimal program for one-way transfers (``minimal'').
We used minimal to establish a performance baseline for comparison with the other programs we evaluated.
We also used minimal to record exactly which packets were dropped in each transfer.
Next, we developed a more full-featured program for one-way file transfers (``pydiode'').
pydiode includes mitigations for packet loss and can detect transfer errors.

\subsubsection*{Developing a Minimal Program}
To establish a performance baseline, we developed a minimal program for one-way transfers.
Our non-functional design requirements were that the software: be cross-platform, be understandable by the widest audience possible, minimize the number of lines of code, and avoid dependencies.
Thus, we developed the software in Python.
When sending, minimal reads data from standard input, and sends the data in UDP packets to the receiver.
When receiving, minimal listens for UDP packets, writes their payloads to standard output, and ends the transfer after 200~ms elapse without receiving another packet.
We initially tested using a 100~ms timeout, but this caused some transfers to exit prematurely on macOS due to timer coalescing~\cite{PowerEfficiency}.
We include configuration options to limit the maximum bitrate, adjust the maximum size of packets, request a different receive buffer size, and to change how data is written to standard output by the receiver.
We observed that netcat is single-threaded, whereas UDPcast and lidi are multithreaded, and we wanted to measure the impact of multithreading on performance.
Thus, we included options to control how the receiver writes to standard output: after the transfer completes, using a helper thread, using a thread orchestrated by asyncio, and simply writing using the main thread.
Finally, we included a feature which gave us detailed insight into exactly which packets are dropped in each transfer.
When the packet details feature is enabled, both the sender and receiver save each packet in memory until the transfer completes.
Afterwards, they write CSV files containing the SHA-256 digests of each packet.
By comparing the CSV files from the sender and receiver, the indices of each dropped packet can be determined.
Figure~\ref{fig:minimal} in Appendix~\ref{sec:figures} shows the core of our program.
Not counting comments, whitespace, argument parsing, and packet details logging, the core of our program includes just 44 lines of code.

\subsubsection*{Developing pydiode}

\begin{figure*}
\centering
\includegraphics[width=\linewidth]{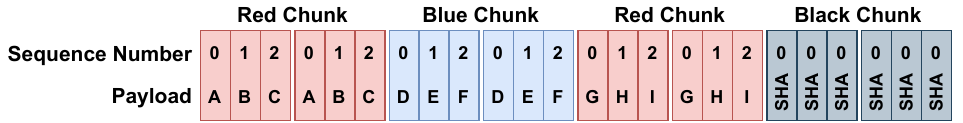}
\caption{pydiode sends packets in ``chunks,'' which alternate between ``red'' and ``blue.''
For resilience against packet loss, each chunk is transmitted a configurable number of times.
The end-of-file is signalled by a ``black'' chunk, which contains a SHA-256 digest of the transmitted data.
This diagram shows three distinct chunks of data.
Each chunk of data consists of three packets, with sequence numbers 0, 1, and 2.
}
\label{fig:pydiode}
\end{figure*}

We used the findings from our minimal program (\S~\ref{sec:minimal_linux_results} and \S~\ref{sec:minimal_macos_results}) to develop pydiode.
pydiode is designed for reliable one-way transfers on Linux and macOS, without needing superuser privileges.
pydiode uses redundancy to mitigate the patterns of packet loss we observed when testing minimal.
pydiode sends chunks of data a configurable number of times, where each chunk is composed of a configurable number of packets.
The receiver uses each packet's sequence number to reassemble the original chunk of data, even if packets arrive out of order.
The receiver waits to output data until all of a chunk's packets have been received.
To distinguish between chunks, pydiode alternates the color of the chunk between ``red'' and ``blue.''
When the data transfer is complete, a ``black'' chunk is sent, indicating the end-of-file.
Each ``black'' packet also contains a SHA-256 digest of the transmitted data, which the receiver compares to its received data to detect transfer failures.
Each packet consists of a seven byte header, the payload, and possibly padding to ensure uniform packet sizes.
The header includes the chunk color, the number of packets in the chunk, the sequence number of the packet, and the payload length.
Figure~\ref{fig:pydiode} depicts a series of packets sent by pydiode.

On Linux and macOS, we observed rare, short clusters of packet loss (up to 80 packets) at unpredictable locations within transfers.
To mitigate packet loss, pydiode defaults to a chunk length of 100 packets, and sends each chunk twice: this makes pydiode resilient to packet loss of up to 100 sequential packets.
Also, on both Linux and macOS we identified configurations for which lower bitrates were less reliable than higher bitrates.
Thus, pydiode uses retransmission and padding to send data at the configured bitrate, even if pydiode's input stream supplies data at a lower rate.
pydiode enforces a maximum bitrate by sleeping after sending each packet, and pydiode uses a helper thread to output received data.
On Linux, pydiode sends 65,507 byte payloads.
On macOS, pydiode sends 1,472 byte payloads, since macOS does not allow sending larger broadcast packets.
pydiode requests an 8.4~MB receive buffer size, which is allowed on macOS, but is ignored on Linux.

%% file: 5-results.tex
\section{Results}
\label{sec:results}

First, we use our minimal program to characterize packet loss on Linux (\S~\ref{sec:minimal_linux_results}) and macOS (\S~\ref{sec:minimal_macos_results}).
Next, we evaluate netcat (\S~\ref{sec:netcat_results}), UDPcast (\S~\ref{sec:udpcast_results}), and lidi (\S~\ref{sec:lidi_results}) on Linux.
Finally, we evaluate pydiode on Linux (\S~\ref{sec:pydiode_linux_results}) and macOS (\S~\ref{sec:pydiode_macos_results}).

\subsection{Performance Baseline on Linux}
\label{sec:minimal_linux_results}
We conducted several experiments with our minimal program on Linux.
First, we varied the maximum UDP payload size, using two different receive buffer sizes (Table~\ref{table:minimal_linux_small} and Table~\ref{table:minimal_linux_big}).
Next, we varied the receiver's write method (Table~\ref{table:minimal_linux_write}).
In all experiments, we varied the sender's rate limit, and we used minimal's packet details feature to determine exactly which packets were dropped.
Since the packet details feature takes extra time to write CSV files to disk, the durations in our tables are from 1,000 trials run without packet details.
Our minimal program does not include any forward error correction or redundancy, so transfers will fail if even a single packet is dropped.
Note that Linux blocks outgoing datagrams until it is ready to send them, so minimal's unlimited configuration is limited by the bandwidth of the underlying network hardware.

\begin{table*}
\begin{subtable}{0.49\linewidth}
\centering
\begin{adjustbox}{max width=\linewidth}
\begin{tabular}{| r | r | r | r |}
\multicolumn{4}{c}{\textbf{minimal on Linux, 213 kB Receive Buffer}} \\ \hline
\small \textbf{Max Bitrate} &
\small \textbf{Max Payload} &
\small \textbf{Succeeded} &
\small \textbf{Duration} \\
\hline
500 Mbit/s &  1472 & 100.00\% & 7.33 s \\ \hline
750 Mbit/s &  1472 & 100.00\% & 6.39 s \\ \hline
  1 Gbit/s &  1472 &  99.99\% & 5.70 s \\ \hline
 unlimited &  1472 &   9.85\% & 1.32 s \\ \hline \hline
500 Mbit/s &  9216 & 100.00\% & 3.03 s \\ \hline
750 Mbit/s &  9216 &  99.98\% & 2.39 s \\ \hline
  1 Gbit/s &  9216 &  99.97\% & 2.03 s \\ \hline
 unlimited &  9216 &  99.36\% & 1.32 s \\ \hline \hline
\rowcolor{green!25} 500 Mbit/s & 65507 & 100.00\% & 2.45 s \\ \hline
750 Mbit/s & 65507 &  99.99\% & 1.74 s \\ \hline
  1 Gbit/s & 65507 &  99.94\% & 1.40 s \\ \hline
 unlimited & 65507 &  99.95\% & 1.31 s \\ \hline
\end{tabular}
\end{adjustbox}
\caption{
On Linux, programs are limited to a 213 kB receive buffer by default.
}
\label{table:minimal_linux_small}
\end{subtable}
\hfill
\begin{subtable}{0.49\linewidth}
\centering
\begin{adjustbox}{max width=\linewidth}
\begin{tabular}{| r | r | r | r |}
\multicolumn{4}{c}{\textbf{minimal on Linux, 8.4 MB Receive Buffer}} \\ \hline
\small \textbf{Max Bitrate} &
\small \textbf{Max Payload} &
\small \textbf{Succeeded} &
\small \textbf{Duration} \\
\hline
\rowcolor{yellow!25} 500 Mbit/s &  1472 & 100.00\% & 7.33 s \\ \hline
\rowcolor{yellow!25} 750 Mbit/s &  1472 & 100.00\% & 6.39 s \\ \hline
\rowcolor{yellow!25}   1 Gbit/s &  1472 & 100.00\% & 5.69 s \\ \hline
\rowcolor{yellow!25}  unlimited &  1472 &  99.99\% & 1.32 s \\ \hline \hline
\rowcolor{yellow!25} 500 Mbit/s &  9216 & 100.00\% & 3.03 s \\ \hline
\rowcolor{yellow!25} 750 Mbit/s &  9216 & 100.00\% & 2.39 s \\ \hline
\rowcolor{yellow!25}   1 Gbit/s &  9216 &  99.99\% & 2.03 s \\ \hline
\rowcolor{yellow!25}  unlimited &  9216 &  99.99\% & 1.32 s \\ \hline \hline
\rowcolor{yellow!25} 500 Mbit/s & 65507 & 100.00\% & 2.45 s \\ \hline
\rowcolor{yellow!25} 750 Mbit/s & 65507 & 100.00\% & 1.74 s \\ \hline
\rowcolor{yellow!25}   1 Gbit/s & 65507 & 100.00\% & 1.40 s \\ \hline
\rowcolor{yellow!25}  unlimited & 65507 &  99.99\% & 1.31 s \\ \hline
\end{tabular}
\end{adjustbox}
\caption{
For comparison with macOS, we used superuser privileges to request an 8.4 MB receive buffer.
}
\label{table:minimal_linux_big}
\end{subtable}%
\hfill

\vspace{2mm}

\begin{subtable}{0.49\linewidth}
\centering
\begin{adjustbox}{max width=\linewidth}
\begin{tabular}{| r | r | r | r |}
\multicolumn{4}{c}{\textbf{minimal on macOS, 213 kB Receive Buffer}} \\ \hline
\small \textbf{Max Bitrate} &
\small \textbf{Max Payload} &
\small \textbf{Succeeded} &
\small \textbf{Duration} \\
\hline
500 Mbit/s &  1472 &  0.35\% & 3.38 s \\ \hline
750 Mbit/s &  1472 &  0.00\% & 2.35 s \\ \hline
  1 Gbit/s &  1472 &  0.00\% & 1.84 s \\ \hline \hline
\rowcolor{yellow!25} 500 Mbit/s &  9216 & 97.99\% & 3.11 s \\ \hline
\rowcolor{yellow!25} 750 Mbit/s &  9216 & 99.97\% & 2.17 s \\ \hline
\rowcolor{yellow!25}   1 Gbit/s &  9216 & 99.98\% & 1.86 s \\ \hline \hline
\rowcolor{yellow!25} 500 Mbit/s & 65507 & 98.22\% & 3.20 s \\ \hline
\rowcolor{yellow!25} 750 Mbit/s & 65507 & 98.50\% & 2.20 s \\ \hline
\rowcolor{yellow!25}   1 Gbit/s & 65507 & 98.90\% & 1.66 s \\ \hline
\end{tabular}
\end{adjustbox}
\caption{
For comparison with Linux, we requested a 213 kB receive buffer.
}
\label{table:minimal_macos_small}
\end{subtable}
\hfill
\begin{subtable}{0.49\linewidth}
\centering
\begin{adjustbox}{max width=\linewidth}
\begin{tabular}{| r | r | r | r |}
\multicolumn{4}{c}{\textbf{minimal on macOS, 8.4 MB Receive Buffer}} \\ \hline
\small \textbf{Max Bitrate} &
\small \textbf{Max Payload} &
\small \textbf{Succeeded} &
\small \textbf{Duration} \\
\hline
500 Mbit/s &  1472 & 100.00\% & 3.38 s \\ \hline
750 Mbit/s &  1472 & 100.00\% & 2.35 s \\ \hline
\rowcolor{green!25}   1 Gbit/s &  1472 & 100.00\% & 1.84 s \\ \hline \hline
\rowcolor{yellow!25} 500 Mbit/s &  9216 & 100.00\% & 3.11 s \\ \hline
\rowcolor{yellow!25} 750 Mbit/s &  9216 & 100.00\% & 2.17 s \\ \hline
\rowcolor{yellow!25}   1 Gbit/s &  9216 & 100.00\% & 1.86 s \\ \hline \hline
\rowcolor{yellow!25} 500 Mbit/s & 65507 & 100.00\% & 3.20 s \\ \hline
\rowcolor{yellow!25} 750 Mbit/s & 65507 & 100.00\% & 2.20 s \\ \hline
\rowcolor{yellow!25}   1 Gbit/s & 65507 &  99.97\% & 1.66 s \\ \hline
\end{tabular}
\end{adjustbox}
\caption{
On macOS, programs can request up to an 8.4 MB receive buffer without using superuser privileges.
}
\label{table:minimal_macos_big}
\end{subtable}
\caption{minimal results on Linux and macOS, showing the effect of payload size and receive buffer size.
We always used the helper write method.
For each configuration, we performed 10,000 trials transferring 1~Gbit of data.
The fastest reliable configurations which do not require superuser privileges are colored green.
Configurations which require superuser privileges are colored yellow.}
\label{table:minimal_results}
\end{table*}

\begin{table}
\centering
\begin{adjustbox}{max width=\linewidth}
\begin{tabular}{| r | r | r | r |}
\multicolumn{4}{c}{\textbf{minimal on Linux, Comparing Write Method}} \\ \hline
\small \textbf{Max Bitrate} &
\small \textbf{Write Method} &
\small \textbf{Succeeded} &
\small \textbf{Duration} \\
\hline
500 Mbit/s &  helper & 100.00\% & 2.45 s \\ \hline
750 Mbit/s &  helper & 100.00\% & 1.74 s \\ \hline
  1 Gbit/s &  helper &  99.92\% & 1.40 s \\ \hline
 unlimited &  helper &  99.81\% & 1.31 s \\ \hline \hline
500 Mbit/s &   defer & 100.00\% & 2.49 s \\ \hline
750 Mbit/s &   defer & 100.00\% & 1.78 s \\ \hline
  1 Gbit/s &   defer &  99.99\% & 1.44 s \\ \hline
 unlimited &   defer &  99.94\% & 1.35 s \\ \hline \hline
500 Mbit/s & asyncio & 100.00\% & 2.52 s \\ \hline
750 Mbit/s & asyncio &  99.99\% & 1.89 s \\ \hline
  1 Gbit/s & asyncio &  99.98\% & 1.52 s \\ \hline
 unlimited & asyncio &  99.91\% & 1.39 s \\ \hline \hline
500 Mbit/s &    main &  32.62\% & 2.44 s \\ \hline
750 Mbit/s &    main &  75.29\% & 1.74 s \\ \hline
  1 Gbit/s &    main &  99.11\% & 1.40 s \\ \hline
 unlimited &    main &  99.61\% & 1.31 s \\ \hline
\end{tabular}
\end{adjustbox}
\caption{
On Linux, we varied the receiving program's write method.
We used 65,507 byte UDP payloads and a 213 kB receive buffer, the maximum without using superuser priviledges.
For each configuration, we performed 10,000 trials transferring 1 Gbit of data.
}
\label{table:minimal_linux_write}
\end{table}

Our results show that increasing the receive buffer size from 213~kB to 8.4~MB made transfers significantly more reliable.
With an 8.4~MB receive buffer, only one transfer failed for four different configurations.
Unfortunately, increasing the receive buffer requires superuser privileges.
However, we identified multiple reliable configurations which do not require superuser privileges.
\textbf{The fastest of these reliable configurations used a helper thread to write to standard output, 65,507 byte payloads, and limited the bitrate to 500 Mbit/sec.}
Notably, we did not identify any reliable configurations when the main thread was used for writing.
Inspecting our instrumentation data, we see that \textbf{almost all transfer failures are accompanied by a nonzero number of RcvbufErrors, indicating that the receiving UDP buffer overflowed.}
Across 400,000 trials and 18,484 failures, only eight of the failures were not accompanied by RcvbufErrors.
If a program uses a single thread, each write to standard output blocks the program from reading from the UDP buffer for a short period of time, during which time the buffer may overflow, causing incoming packets to be discarded.
Thus, \textbf{the receiving program should use separate threads for reading from the network and writing to standard output.}
Next, observe that \textbf{transfers were more reliable when larger packets were sent.}
When using larger packets, fewer packets are needed.
Since minimal handles each packet as it arrives, fewer packets results in fewer system calls and greater responsiveness.
This pattern is clearest when using an unlimited bitrate: when rate-limiting while sending smaller packets, more sleeps are performed, and due to operating system scheduling data is sent even more slowly than requested.
Finally, notice that for the helper, defer, and asyncio write methods, transfers are more reliable at lower bitrates.
In contrast, the main write method performs much worse at 500 Mbit/sec and 750 Mbit/sec than at higher bitrates.
Thus, \textbf{lower-bitrate transfers are not necessarily more reliable than higher-bitrate transfers.}

Next, we analyzed the packet details CSV files from the sender and receiver.
By comparing the CSV files, we determined the indices of each dropped packet.
First, we compared the number of RcvbufErrors to the number of packets that didn't reach the receiving program.
In all but eight trials, the number of RcvbufErrors exactly matched the number of dropped packets.
Considering all trials, \textbf{99.99\% of packet loss was due to RcvbufErrors, confirming that most packet loss is due to the receiving program being insufficiently responsive.}
The unexplained packet loss occurs lower in the network stack, possibly due to the Linux kernel itself dropping packets, or hardware-level errors.
Unexplained packet loss occurred at a rate of approximately one in $5 \times 10^7$ packets.
However, the packet loss was not randomly distributed: in the trials with unexplained packet loss, between one and 109 packets were dropped, and when multiple packets were dropped, they were either adjacent or in close proximity.

The fastest reliable configurations we identified used the helper write method, 65,507 byte payloads, and a bitrate limit of 500~Mbit/sec.
Faster transfers could be acheived if packet loss was mitigated using redundancy or forward error correction, rather than a rate limit.
Thus, we analyzed exactly which packets were dropped in the corresponding unlimited bitrate transfers.
In total, 25 of 30,000 trials failed.
Of these failed trials, 24 trials had packet loss due to RcvbufErrors, and one trial had packet loss due to unexplained reasons.
Between one and six packets were dropped, and when multiple packets were dropped, they were either adjacent or in close proximity.
The maximum difference between the indices of the first and last dropped packets was eight.
Thus, both the receiving UDP buffer overflowing and issues lower in the network stack can cause multiple packets to be dropped in succession.
\textbf{Since packet loss occurs in clusters, error correction should include redundant information some distance from the data it is protecting.}

\subsection{Performance Baseline on macOS}
\label{sec:minimal_macos_results}
Our preliminary testing on macOS showed that with an unlimited transfer rate, all transfers failed, and transfers finished far too quickly.
This is because the sendto function doesn't block on macOS, a behavior it inherits from FreeBSD~\cite{errorReceived,sendtoMacOS}.
Instead, outgoing datagrams are discarded if the network stack is not ready to send them.
Thus, our code must rate-limit calls to sendto.
Another challenge is that although macOS can receive datagrams of any size, macOS limits the size of outgoing datagrams.
By default, outgoing broadcast payloads are limited to 1,472 bytes.
Packets directed to unicast addresses can have payloads of up to 9,216 bytes, but this requires creating a manual ARP entry using superuser privileges.
Using superuser privileges, it is also possible to enable unicast packets with payloads up to 65,507 bytes, the maximum possible for IPv4 UDP packets.

Table~\ref{table:minimal_results} compares the results from our minimal program on Linux and macOS.
Programs on macOS can request receive buffers up to 8.4~MB, whereas on Linux superuser priviledges are required to exceed the limit of 213~kB.
For comparison with Linux, we tested using both 213~kB and 8.4~MB receive buffers (Table~\ref{table:minimal_macos_small} and Table~\ref{table:minimal_macos_big}, respectively).
We varied the maximum UDP payload size and bitrate, and all configurations used the helper write method.

Consistent with our Linux results, we see a strong association between RcvbufErrors and transfer failures: only one transfer failed without an accompanying RcvbufError, and only one packet was dropped in that transfer.
Since macOS does not support network namespaces, our network statistics are affected by other traffic on the device, making it difficult to count exactly how much packet loss was due to factors other than RcvbufErrors.
However, all the transfers which succeeded had zero RcvbufErrors, and the transfers which failed had an average of 1,379 RcvbufErrors.
As on Linux, lower bitrates are sometimes \textit{less reliable} than faster bitrates.
\textbf{Different than our Linux results, 65,507 byte payloads can be less reliable than smaller payloads.}
Only three transfers failed when using an 8.4~MB buffer, and all had 65,507 byte payloads and a 1~Gbit/sec bitrate.
Examining exactly which packets were dropped in these transfers, we saw one cluster of packet loss in each failed transfer, with between 13 and 80 immediately adjacent packets dropped.
The number of dropped packets exactly matches the number of RcvbufErrors recorded for those transfers.
Our results also show that \textbf{increasing the receive buffer size significantly improves reliability.}
This effect is especially pronounced when using 1,472 byte payloads and a 1~Gbit/sec bitrate: all transfers succeeded with an 8.4 MB buffer, whereas all transfers failed with a 213 kB buffer.
\textbf{Despite macOS limiting the size of outgoing broadcast packets to 1472 bytes, reliable transfers can be achieved by requesting a larger receive buffer.}

\subsection{Evaluating netcat on Linux}
\label{sec:netcat_results}
netcat does not offer any configurable options which would increase the reliability of transfers, so we simply performed 10,000 trials transferring 1~Gbit of data.
93.89\% of trials completed successfully, in an average of 2.07 seconds.
Since netcat does not exit until one second has elapsed since it last received a packet, the data itself was transferred in 1.07 seconds.
Transfers failed exactly when there were a nonzero number of RcvbufErrors, indicating that the receiving UDP buffer overflowed.

netcat is written in C, which could offer an advantage relative to minimal's Python code.
However, netcat often performed worse than our minimal program.
Since netcat is single-threaded, it is particularly interesting that netcat performed worse than minimal's single-threaded (i.e., main write method), unlimited bitrate, 65,507 byte payload configuration: 93.89\% vs 99.61\% of transfers succeeded for netcat and minimal, respectively (Table~\ref{table:minimal_linux_write}).
Whereas minimal was configured to send packets with 65,507 byte payloads, tcpdump shows that netcat sends packets with 16,384 byte payloads.
Furthermore, strace shows that both netcat and minimal trigger write syscalls for each packet they receive.
These findings are consistent with our minimal results showing that using larger payloads improves reliability on Linux (Table~\ref{table:minimal_linux_small}).
minimal's superior performance relative to netcat is likely due to processing fewer packets, or writing to standard output fewer times.

\subsection{Evaluating UDPcast on Linux}
\label{sec:udpcast_results}

Table~\ref{table:udpcast_results} shows the results from testing UDPcast.
We measured the effect of rate limiting and forward error correction (FEC) on transfer reliability.
First, notice that without FEC, transfers have high failure rates.
Second, note that some FEC configurations were more reliable than others.
In particular, when using the 8x8/128 and 8x16/128 FEC configurations, none of the tested bitrates were reliable.
In contrast, when using the 8x32/128 configuration at 500 Mbit/sec or 750 Mbit/sec bitrates, all transfers succeeded.
These configurations used 8, 16, and 32 FEC packets per stripe, respectively.
When there were no RcvbufErrors, all transfers succeeded, but transfers sometimes failed when there were RcvbufErrors.
For example, when using the 8x8/128 FEC configuration, 49\% of transfers had nonzero RcvbufErrors. %
Of those transfers with nonzero RcvbufErrors,
4\% failed %
and 96\% succeeded, %
with a median of 238 and 7 RcvbufErrors, respectively.
As expected, FEC has difficulty mitigating large numbers of dropped packets.

UDPcast is written in C and is multithreaded, so it is suprising that our minimal program outperforms UDPcast.
However, UDPcast sends packets with 1,472 byte payloads, and it does not support sending packets with larger payloads~\cite{udpcastCMD}.
The results from our minimal program suggest that UDPcast would be more reliable if it was modified to send larger packets.
Finally, in one instance the receiving program failed to exit after running for multiple hours.
This trial was configured with a 750 Mbit/sec bitrate and 8x8/128 FEC.
In Table~\ref{table:udpcast_results}, this trial is represented as a failure with an undefined duration.
After we manually reran the sending program, the receiving program exited normally (i.e., with zero as its exit code).
Thus, we infer that the receiving program did not receive the hello packet from the first run of the sending program~\cite{udpcastCMD}.
We attempted an experiment in which UDPcast was configured to use an additional transmission of the hello packet, and we encountered the same problem, so it is unclear how to solve the issue.

\begin{table}
\centering
\begin{adjustbox}{max width=\linewidth}
\begin{tabular}{| r | r | r | r |}
\multicolumn{4}{c}{\textbf{UDPcast on Linux}} \\ \hline
\small \textbf{Max Bitrate} &
\small \textbf{FEC} &
\small \textbf{Succeeded} &
\small \textbf{Duration} \\
\hline
500 Mbit/s &     None &  82.62\% & 2.61 s \\ \hline
750 Mbit/s &     None &  76.99\% & 1.90 s \\ \hline
  1 Gbit/s &     None &  77.97\% & 1.60 s \\ \hline
\rowcolor{red!25} unlimited &     None &  78.79\% & 1.61 s \\ \hline \hline
500 Mbit/s &  8x8/128 &  99.78\% & 4.03 s \\ \hline
750 Mbit/s &  8x8/128 &  99.05\% & 3.19 s \\ \hline
  1 Gbit/s &  8x8/128 &  96.67\% & 2.78 s \\ \hline
   unlimited &  8x8/128 &  96.63\% & 2.79 s \\ \hline \hline
500 Mbit/s & 8x16/128 &  99.99\% & 4.45 s \\ \hline
750 Mbit/s & 8x16/128 &  99.87\% & 3.41 s \\ \hline
  1 Gbit/s & 8x16/128 &  99.61\% & 2.98 s \\ \hline
   unlimited & 8x16/128 &  99.63\% & 2.99 s \\ \hline \hline
500 Mbit/s & 8x32/128 & 100.00\% & 5.09 s \\ \hline
\rowcolor{green!25} 750 Mbit/s & 8x32/128 & 100.00\% & 3.98 s \\ \hline
  1 Gbit/s & 8x32/128 &  99.88\% & 3.52 s \\ \hline
   unlimited & 8x32/128 &  99.91\% & 3.52 s \\ \hline
\end{tabular}
\end{adjustbox}
\caption{We tested transferring 1 Gbit of data using UDPcast on Linux.
For each configuration, we performed 10,000 trials.
The fastest reliable configuration we identified is colored green.
The default, unreliable configuration is colored red.}
\label{table:udpcast_results}
\end{table}

\begin{table}
\centering
\begin{adjustbox}{max width=\linewidth}
\begin{tabular}{| r | r | r | r | r |}
\multicolumn{4}{c}{\textbf{lidi on Linux}} \\ \hline
\small \textbf{Max Bitrate} &
\small \textbf{FEC} &
\small \textbf{Succeeded} &
\small \textbf{Duration} \\
\hline
\rowcolor{red!25}   unlimited &   2\% &  99.29\% & 1.11 s \\ \hline
                    unlimited &  10\% &  99.53\% & 1.19 s \\ \hline
                    unlimited &  25\% &  99.95\% & 1.35 s \\ \hline
\rowcolor{green!25} unlimited &  50\% & 100.00\% & 1.61 s \\ \hline
                    unlimited & 100\% & 100.00\% & 2.14 s \\ \hline
\end{tabular}
\end{adjustbox}
\caption{We tested transferring 1 Gbit of data using lidi on Linux.
For each configuration, we performed 10,000 trials.
The fastest reliable configuration we identified is colored green.
The default, unreliable configuration is colored red.}
\label{table:lidi_results}
\end{table}

\subsection{Evaluating lidi on Linux}
\label{sec:lidi_results}
Table~\ref{table:lidi_results} shows the results from testing lidi.
lidi's transfer rate is not configurable, so we only measured the effect of FEC on reliability.
We observed that when lidi's transfers failed, the receiving program never exited.
Thus, we used to the \texttt{timeout} command to terminate the receiving program after 60 seconds, and we counted these transfers as failures with undefined durations.
Transfers were reliable when we increased the amount of FEC data from the default of 2\% to at least 50\%.

Similar to UDPcast, when there were no RcvbufErrors, all transfers succeeded, but transfers sometimes failed when there were RcvbufErrors.
For example, when using the 2\% FEC setting, 79 transfers had nonzero RcvbufErrors.
Of those 2\% FEC transfers with nonzero RcvbufErrors,
71 failed and only eight succeeded,
with a median of 57 and 2.5 RcvbufErrors, respectively.
When using the 50\% FEC setting, 93 transfers had nonzero RcvbufErrors.
Of those 50\% FEC transfers with nonzero RcvbufErrors,
all succeeded, and there were a median of 52 RcvbufErrors.

Based on our minimal program's behavior, we expect that lidi's packet loss occurs in clusters, meaning that many packets are lost from the same block of data.
Thus, a large amount of FEC data is required to mitigate large numbers of dropped packets.
Our minimal program showed that transfers encounter more RcvbufErrors when small payloads are sent.
Using tcpdump, we observe that lidi sends packets with a maximum payload of 1,468 bytes.
Similar to our minimal program, lidi might encounter fewer RcvbufErrors if it sent larger payloads.

\subsection{Evaluating pydiode on Linux}
\label{sec:pydiode_linux_results}
We applied the findings from our minimal program to develop a more robust program, pydiode (\S~\ref{sec:sw_dev}).
On Linux, we tested pydiode using 65,507 byte payloads sent to a broadcast IP address, a configuration which does not require superuser privileges.
Table~\ref{table:pydiode_linux_results} shows the effect of rate limiting and redundancy on transfer reliability.
\textbf{No transfers failed when redundant transmissions were enabled,} showing that redundancy mitigates the clusters of packet loss we measured using our minimal program.
Also, notice that pydiode's single-redundancy unlimited transfers completed faster than similarly configured minimal transfers: this is because pydiode exits after receiving an end-of-file packet, whereas minimal waits until 200~ms elapse without receiving another packet.

\begin{table*}
\begin{subtable}{0.49\linewidth}
\centering
\begin{adjustbox}{max width=\linewidth}
\begin{tabular}{| r | r | r | r |}
\multicolumn{4}{c}{\textbf{pydiode on Linux}} \\ \hline
\small \textbf{Max Bitrate} &
\small \textbf{Redundancy} &
\small \textbf{Succeeded} &
\small \textbf{Duration} \\
\hline
500 Mbit/s & 1 & 100.00\% & 2.54 s \\ \hline
750 Mbit/s & 1 &  99.89\% & 1.74 s \\ \hline
  1 Gbit/s & 1 &  99.74\% & 1.35 s \\ \hline
   unlimited & 1 &  99.62\% & 1.21 s \\ \hline \hline
500 Mbit/s & 2 & 100.00\% & 5.02 s \\ \hline
750 Mbit/s & 2 & 100.00\% & 3.40 s \\ \hline
\rowcolor{blue!25} 1 Gbit/s & 2 & 100.00\% & 2.64 s \\ \hline
\rowcolor{green!25} unlimited & 2 & 100.00\% & 2.35 s \\ \hline
\end{tabular}
\end{adjustbox}
\caption{On Linux, pydiode uses 65,507 byte payloads and is limited to a 213 kB receive buffer.}
\label{table:pydiode_linux_results}
\end{subtable}
\hfill
\begin{subtable}{0.49\linewidth}
\centering
\begin{adjustbox}{max width=\linewidth}
\begin{tabular}{| r | r | r | r |}
\multicolumn{4}{c}{\textbf{pydiode on macOS}} \\ \hline
\small \textbf{Max Bitrate} &
\small \textbf{Redundancy} &
\small \textbf{Succeeded} &
\small \textbf{Duration} \\
\hline
500 Mbit/s & 1 & 100.00\% & 3.21 s \\ \hline
750 Mbit/s & 1 & 100.00\% & 2.18 s \\ \hline
\rowcolor{green!25} 1 Gbit/s & 1 & 100.00\% & 1.65 s \\ \hline \hline
500 Mbit/s & 2 & 100.00\% & 6.32 s \\ \hline
750 Mbit/s & 2 & 100.00\% & 4.25 s \\ \hline
\rowcolor{blue!25} 1 Gbit/s & 2 & 100.00\% & 3.18 s \\ \hline
\end{tabular}
\end{adjustbox}
\caption{On macOS, pydiode uses 1,472 byte payloads and an 8.4 MB receive buffer.}
\label{table:pydiode_macos_results}
\end{subtable}
\caption{pydiode results on Linux and macOS.
For each configuration, we performed 10,000 trials transferring 1 Gbit of data.
On each platform, the fastest reliable configuration we identified is colored green.
The default configurations are also reliable but slower, and are colored blue.
}
\label{table:pydiode_results}
\end{table*}

\subsection{Evaluating pydiode on macOS}
\label{sec:pydiode_macos_results}
On macOS, we tested pydiode using 1,472 byte payloads sent to a broadcast IP address, a configuration which does not require superuser privileges.
Table~\ref{table:pydiode_macos_results} shows the effect of rate limiting and redundancy on transfer reliability.
Consistent with our minimal results, \textbf{all redundant and non-redundant transfers completed successfully.}

%% file: 6-limitations.tex
\section{Limitations}
\label{sec:limitations}

There are several limitations to consider when interpreting our results.
First, when testing netcat, UDPcast, lidi, and pydiode, we only tested configurations which did not require superuser privileges.
In particular, we did not increase network buffer sizes beyond the limits imposed by macOS and Linux, and we did not modify process priority.
Configurations which require superuser privileges are incompatible with certain contexts (e.g., mobile devices).
To increase the generalizability of our findings, we focused on configurations available without superuser privileges.
Consistent with prior work, our minimal results (Table~\ref{table:minimal_results}) show that larger network buffer sizes reduce packet loss~\cite{stevensImplicationOpticalData1999,pietre-cambacedesDeconstructionIndustrialControl2009,linResearchPacketLoss2013}.
Second, we only tested using two platforms: Intel NUC 12 mini PCs running Ubuntu Desktop 24.04.3 LTS, and M1 Mac minis running macOS Sequoia 15.7.4.
Our results show that transfer reliability varies by platform, and we did not test using mobile devices and Microsoft Windows.
However, mobile devices overwhelmingly run Android and iOS, which share network stacks with Linux and macOS, respectively.
Thus, our results will serve as a useful point of comparison when testing on other platforms.

Our results show that commodity hardware and open source software can be used to implement reliable one-way data transfers.
However, real-world deployments will also depend on the reliability of other components.
During our testing, we discovered rare but persistent reliability issues with the underlying platforms.
When testing with the PCs, the SSH connections between the machines which we used to automate testing occasionally terminated unexpectedly.
The issue was isolated to the secondary PCI Ethernet port, and never occurred after we switched to a USB-C Ethernet port.
We encountered no issues using the PCs' primary built-in Ethernet ports for one-way traffic.
When testing using the M1 Mac minis, we encountered repeated kernel panics, suggesting that we triggered a kernel bug.
Most panics referenced the built-in Ethernet port (AppleT810xPCIePort).
We encountered the panics on three different M1 Mac minis, on both the sender and receiver, and when using both macOS Tahoe 26.2 and macOS Sequoia 15.7.4.
Panics even occurred when no superuser commands had been used.
When we encountered a kernel panic, we restarted the experiment we were running from the beginning.
Although there is an element of randomness, we can reproduce the kernel panics: when repeatedly sending 1~Tbit streams of data using pydiode, we encountered kernel panics after 35, 48, and 71 hours of testing.
We anticipate these kernel panics would be an obstacle to deploying Mac minis in industrial contexts, but would not be an issue for end-users.

%% file: 7-discussion.tex
\section{Discussion}
\label{sec:discussion}

Data diodes are physically limited to only transfer data in one direction, thereby offering a physical defense against sophisticated malware.
Although commercially available data diodes are expensive, data diodes can also be constructed from commodity network hardware (\S~\ref{sec:hardware}).
However, specialized software is needed to send data through a data diode reliably: the receiving program cannot request retransmission of dropped packets, so packet loss must be minimized and mitigated.
In this section, we summarize our results to answer our research questions.
First, we explain the cause of packet loss in data diodes, and ways to mitigate packet loss (\S~\ref{sec:receiving_program}).
Next, we describe the relative performance of the data transfer software we evaluated (\S~\ref{sec:comparison}).
Finally, we offer advice to software developers based on our own experience (\S~\ref{sec:advice}).

\subsection{Packet Loss Is Caused by an Insufficiently Responsive Receiving Program}
\label{sec:receiving_program}
Packet loss between the sending and receiving programs interferes with reliable transmission of data.
But why is there packet loss between two devices which have a direct physical connection?
We transferred terabytes of data through our data diode using our minimal program, and we found that 99.99\% of packet loss was caused by receiving program unresponsiveness (\S~\ref{sec:minimal_linux_results}).
When the receiving program processes incoming packets too slowly, its UDP buffer overflows, causing clusters of packet loss.
Rare packet loss from lower in the network stack also occurred in clusters.
To mitigate clusters of packet loss, redundant information should not be sent in adjacent packets.

We discovered several ways to minimize packet loss without using superuser privileges.
First, packet loss can be minimized by using a separate helper thread to write out received data.
Second, programs on Linux can send packets with larger payloads.
Third, programs on macOS can request a larger receive buffer size.
We also made several counterintuitive discoveries.
First, on both Linux and macOS, we observed cases where transferring data more slowly \textit{decreased} transfer reliability.
This contradicts recommendations from prior work~\cite{pietre-cambacedesDeconstructionIndustrialControl2009,linResearchPacketLoss2013}.
If the sending program reads from a stream of data, its input bandwidth may vary over time.
This suggests that the sending program should use retransmission and padding to maintain a consistent output bandwidth for more predictable performance.
Second, although we found that sending larger packets increased reliability on Linux, sending larger packets on macOS \textit{decreased} reliability in some cases.
Thus, the optimal payload size is platform-dependent.
We incorporated our findings into the development of pydiode, a cross-platform program for reliably transferring information through data diodes (\S~\ref{sec:sw_dev}).

\subsection{When Properly Configured, Multiple Programs Offer Reliable Transfers}
\label{sec:comparison}
We tested three existing open source programs on Linux: netcat, UDPcast, and lidi.
We found that only 93.89\% of netcat's transfers succeeded (\S~\ref{sec:netcat_results}), and that transfers were also unreliable when using UDPcast (\S~\ref{sec:udpcast_results}) and lidi (\S~\ref{sec:lidi_results}) in their default configurations.
However, we discovered configurations for UDPcast and lidi for which all transfers succeeded.
We also tested our own programs, minimal and pydiode, on both Linux and macOS.
Similar to minimal, pydiode is written in Python, though it differs in two important ways.
First, pydiode uses redundancy to mitigate the clusters of packet loss we observed on Linux and macOS.
Second, the pydiode receiver detects transfer errors by calculating a SHA-256 digest of the data it receives, and comparing it to a SHA-256 digest calculated by the sending program.

\begin{table}
\centering
\begin{adjustbox}{max width=\linewidth}
\begin{tabular}{| r | r | r |}
\multicolumn{3}{c}{\textbf{Fastest Reliable Configuration}} \\ \hline
\textbf{Platform} &
\textbf{Program} &
\textbf{Duration} \\
\hline
Linux & minimal & 2.45 sec \\ \hline
Linux & pydiode & 2.35 sec \\ \hline
Linux & UDPcast & 3.98 sec \\ \hline
Linux & lidi    & 1.61 sec \\ \hline \hline
macOS & minimal & 1.84 sec \\ \hline
macOS & pydiode & 1.65 sec \\ \hline
\end{tabular}
\end{adjustbox}
\caption{For each program, we identified the configuration with the minimum average transfer duration for which all transfers succeeded. We omitted netcat, because only 93.89\% of netcat's trials completed successfully.
}
\label{table:summary}
\end{table}

Table~\ref{table:summary} shows the fastest reliable configuration for each program.
Unlike UDPcast and lidi, neither of our programs employ forward error correction (FEC).
Furthermore, UDPcast and lidi are written in C and Rust, respectively, which offer performance benefits relative to Python.
Nevertheless, both minimal and pydiode offer faster reliable transfers than UDPcast.
Compared to lidi, minimal and pydiode are only 52\% and 46\% slower, respectively.
Our results show that although FEC can improve reliability, it is neither necessary nor sufficient.
In theory, FEC is more efficient than simple retransmission of data, offering greater error correction at a given bitrate.
In practice, if the receiving program does not read from its UDP network buffer frequently enough, the number of packets dropped can exceed what FEC can mitigate.

When choosing between these programs, there are several factors beyond data transfer speed worth considering.
First, some programs may not be available for your operating system.
UDPcast, lidi, and pydiode all run on Linux.
Only pydiode supports macOS.
Currently, only UDPcast supports Windows, though we are working on Windows support for pydiode.
Second, only pydiode supports sending to broadcast addresses.
UDPcast and lidi can only send to unicast addresses, which requires using superuser privileges to create a manual ARP entry.
Third, the codebases have different sizes, which has implications for auditing the code for security vulnerabilities.
pydiode has 1,887 lines of Python code, lidi has 3,592 lines of Rust code, and UDPcast has 7,709 lines of C code.
UDPcast does not require dependencies and pydiode only depends on the Python standard library.
However, lidi requires many direct and indirect Rust dependencies, which include 887,355 additional lines of Rust code.
Finally, lidi uses an open source implementation of RaptorQ codes for forward error correction, but Qualcomm owns many patents related to RaptorQ~\cite{qualcommRaptorQTechnicalOverview2010,RaptorQPatents}.
In contrast, UDPcast and pydiode's underlying technology is public domain.

\subsection{Advice For Software Developers}
\label{sec:advice}
We offer several recommendations to developers seeking to implement software for one-way file transfers.
First, it is important to establish a performance baseline on your platform using a simple program, like our minimal program.
As you implement more complex software with additional features, you should compare it to your performance baseline.
Second, you should incorporate our findings to minimize packet loss (\S~\ref{sec:receiving_program}), considering factors like multithreading, payload size, UDP buffer size, and transfer bitrate.
We recommend focusing on minimizing packet loss before mitigating packet loss.
If you attempt to mitigate packet loss through methods like FEC, you should confirm you are not increasing packet loss by making your program less responsive.
Finally, you should implement your software in a modular fashion.
Stream-based input formats are the most flexible, since they allow sending individual files via file redirection, directories of files using tar, or network traffic using a proxy program.
For example, pydiode's sending program reads from standard input, and its receiving program writes to standard output.
pydiode supports sending directories of files using a process pipeline: Python's tarfile module converts directories to and from streams, and pydiode simply transfers these streams through the data diode.
Similarly, pydiode can transfer network traffic using protocol proxies.
For example, Figure~\ref{fig:mqtt} in Appendix~\ref{sec:figures} shows a proxy for the MQTT protocol, which is widely used in IoT deployments.

%% file: 8-conclusions.tex
\section{Conclusions and Future Work}
\label{sec:conclusions}

Data diodes offer a physical defense against cyberattacks by enforcing the direction of information flow.
If information cannot enter or leave a system, integrity or confidentiality can be ensured, respectively.
Although commercially available data diodes are expensive, our results show that data diodes can be assembled using commodity hardware and open source software for a fraction of the cost.
Our research will unlock new opportunities.
First, security practitioners, educators, and others will benefit from greater access to data diodes.
Security teams can test the suitability of data diodes in their environment with minimal capital investment.
We have also found that data diodes are a useful educational tool in our undergraduate Computer Networks course.
Second, open source solutions support supply chain diversification and avoid vendor lock-in.
Commercially available data diodes use proprietary software, and are not interoperable with each other.
In contrast, open source data diodes are hardware agnostic, and are compatible with fiber-optic network equipment from different vendors.
These benefits are magnified by recent uncertainty in international relations.
Finally, data diodes can be deployed in novel contexts to defend against targeted cyberattacks.
For example, data diodes can be used to harden messaging apps like Signal against spyware~\cite{storyDefendingMessagingApps2026}.

We anticipate several areas for future work.
First, packet loss minimization and mitigation can be implemented more effectively to support faster reliable transfers.
UDPcast and lidi use forward error correction (FEC) to mitigate packet loss, but if their receiving programs were more responsive, they could reduce packet loss significantly.
Similarly, pydiode could be modified to use FEC instead of simple redundancy, allowing more efficient packet loss mitigation.
Second, open source solutions could be compared against commercial products.
Commercial vendors do not publish data on the reliability of their products, so their relative performance is unclear.
Finally, alternative data diode hardware can be developed and evaluated~\cite{OSDD,DYODE,godiode,tfc,NLDOSDD}.

%% file: 9-appendix.tex
\section{Supplementary Figures}
\label{sec:figures}

\begin{figure}[h]
\centering
\begin{Verbatim}[frame=single,fontsize=\small,vspace=0pt]
for i in range(10000):
  generate random data
  for each configuration combination:
    start receiving program
    start sending program
    wait for programs to exit
    compare SHA-256 of sent and received data
\end{Verbatim}
\caption{We tested transferring data through the data diode with different combinations of configurable options. We tested each combination 10,000 times, with randomly generated data each time.}
\label{fig:loop}
\end{figure}

\begin{figure}[h]
\centering
\begin{Verbatim}[frame=single,fontsize=\small,vspace=0pt]
# To send data
nc -u -q 0 -s 10.0.1.2 10.0.1.1 1234 < /tmp/write/random_data

# To receive data
nc -u -w 1 10.0.1.1 -l 1234 > /tmp/random_data
\end{Verbatim}
\caption{Commands to send and receive data using netcat.
The \texttt{-u} option enables UDP instead of TCP.
The \texttt{-s} option specifies the source address.
The \texttt{-w 1} option causes netcat to exit one second after it stops receiving data, meaning that even small transfers take at least one second.
}
\label{fig:netcat_cmds}
\end{figure}

\begin{figure}[h]
\centering
\begin{Verbatim}[frame=single,fontsize=\small,vspace=0pt]
# To send data
udp-sender --interface enp100s0 --mcast-rdv-address 10.0.1.1 \
  --async --rexmit-hello-interval 10 --autostart 1 \
  --max-bitrate 100000000 --fec 8x8/128 < /tmp/write/random_data

# To receive data
udp-receiver --nosync --interface enp100s0 > /tmp/random_data
\end{Verbatim}
\caption{Commands to send and receive data using UDPcast.
The \texttt{-{}-max-bitrate} ``is the raw bitrate, including packet headers, forward error correction, retransmissions, etc. Actual payload bitrate will be lower''~\cite{udpcastCMD}.
The \texttt{-{}-fec} argument specifies how many ``stripes'' each chunk of data is split into, the number of FEC packets included with each stripe, and the number of data packets in each stripe.
The \texttt{-{}-async}, \texttt{-{}-rexmit-hello-interval}, and \texttt{-{}-autostart} arguments are used to run UDPcast in asynchronous mode.
}
\label{fig:udpcast_cmds}
\end{figure}

\begin{figure}[h]
\centering
\begin{Verbatim}[frame=single,fontsize=\small,vspace=0pt]
# To send data
diode-oneshot-send --to 10.0.1.1:1234 --repair 2 < /tmp/write/random_data

# To receive data
timeout 60 diode-oneshot-receive --from 10.0.1.1:1234 --repair 2 > /tmp/random_data
\end{Verbatim}
\caption{Commands to send and receive data using lidi.
The \texttt{-{}-repair} argument controls the amount of repair data, specified as a percent of the original data~\cite{lidiCLI}.
}
\label{fig:lidi_cmds}
\end{figure}

\begin{figure*}
\centering
\begin{Verbatim}[frame=single,fontsize=\footnotesize,vspace=0pt,numbers=left]
import queue
import socket
import sys
import threading
import time

def send(read_ip, read_port, write_ip, max_bitrate, max_payload):
    # To avoid exceeding max_bitrate, take at least this many seconds to send each packet
    target_elapsed = max_payload / max_bitrate * BYTE if max_bitrate else 0
    data = sys.stdin.buffer.read(max_payload)
    with socket.socket(socket.AF_INET, socket.SOCK_DGRAM) as sock:
        sock.setsockopt(socket.SOL_SOCKET, socket.SO_BROADCAST, 1)
        sock.bind((write_ip, 0))  # The OS will choose an available port
        while data:
            start = time.monotonic()
            sock.sendto(data, (read_ip, read_port))
            data = sys.stdin.buffer.read(max_payload)
            if target_elapsed:
                already_elapsed = time.monotonic() - start
                sleep_duration = target_elapsed - already_elapsed
                if sleep_duration > 0:
                    time.sleep(sleep_duration)

def write(packets):
    data = packets.get()
    while data is not None:
        sys.stdout.buffer.write(data)
        data = packets.get()

def receive(read_ip, read_port, rcvbuf_size, timeout):
    packets = queue.Queue()
    received_packets = False
    # Write to STDOUT using a separate thread
    t = threading.Thread(target=write, args=(packets,))
    t.start()
    # Receive packets
    with socket.socket(socket.AF_INET, socket.SOCK_DGRAM) as sock:
        sock.bind((read_ip, read_port))
        sock.setsockopt(socket.SOL_SOCKET, socket.SO_RCVBUF, rcvbuf_size)
        sock.settimeout(timeout)
        while True:
            try:
                # IPv4 UDP packet payloads cannot exceed 65507 bytes
                data, _ = sock.recvfrom(65507)
                received_packets = True
                packets.put(data)
            # Break from the loop if we have received some packets, but
            # timeout seconds have elapsed since receiving the last packet
            except TimeoutError:
                if received_packets:
                    break
    # Indicate there won't be more packets
    packets.put(None)
    t.join()
\end{Verbatim}
\caption{Code from our minimal program. Argument parsing and packet loss logging are omitted. Since writing to standard output using a helper thread offered the best performance, code for the other write methods is omitted.
}
\label{fig:minimal}
\end{figure*}

\begin{figure*}
\centering
\begin{subfigure}{\linewidth}
\begin{Verbatim}[frame=single,fontsize=\footnotesize,vspace=0pt,numbers=left]
import argparse
import csv
import sys
import paho.mqtt.client as mqtt

def write(client, writer, msg):
    writer.writerow({"topic": msg.topic, "payload": msg.payload})
    sys.stdout.flush()

FIELDNAMES = ["topic", "payload"]

def main():
    parser = argparse.ArgumentParser(
        description="Relay MQTT data via STDIN and STDOUT"
    )
    parser.add_argument(
        "mode",
        help="Whether to subscribe or publish MQTT data",
        choices=("subscribe", "publish"),
    )
    parser.add_argument(
        "host",
        help="MQTT broker hostname",
    )
    args = parser.parse_args()
    client = mqtt.Client(mqtt.CallbackAPIVersion.VERSION2)
    if args.mode == "subscribe":
        writer = csv.DictWriter(sys.stdout, fieldnames=FIELDNAMES)
        writer.writeheader()
        client.user_data_set(writer)
        client.on_connect = lambda client, *args: client.subscribe("#")
        client.on_message = write
        client.connect(args.host)
        client.loop_forever()
    else:
        reader = csv.DictReader(sys.stdin, fieldnames=FIELDNAMES)
        client.connect(args.host)
        client.loop_start()
        for row in reader:
            client.publish(row["topic"], row["payload"])
        client.disconnect()
        client.loop_stop()
\end{Verbatim}
\end{subfigure}%

\vspace{2mm}

\begin{subfigure}{\linewidth}
Example usage:
\begin{Verbatim}[frame=single,fontsize=\footnotesize,vspace=0pt]
# To send data from the high-integrity network
python3 relay.py subscribe mqtt.industrial.net | pydiode send 10.0.1.255 10.0.1.2

# To receive data
pydiode receive 10.0.1.255 | python3 relay.py publish mqtt.business.net
\end{Verbatim}
\end{subfigure}
\caption{MQTT protocol proxy. Suppose IoT devices on a high-integrity industrial network publish data to an MQTT broker. By transferring the MQTT broker's data through a data diode, the data can be accessed on the organization's low-integrity business network without exposing the high-integrity network to cyberattacks.}
\label{fig:mqtt}
\end{figure*}

%% file: main.bbl
\begin{thebibliography}{10}

\bibitem{ablonZeroDaysThousands2017}
Lillian Ablon and Andy Bogart.
\newblock Zero {{Days}}, {{Thousands}} of {{Nights}}: {{The Life}} and
  {{Times}} of {{Zero-Day Vulnerabilities}} and {{Their Exploits}}.
\newblock Technical report, RAND Corporation, 2017.
\newblock \url{http://www.rand.org/pubs/research_reports/RR1751.html}.

\bibitem{advisorycommitteeonreactorsafeguardsdigitalinstrumentationandcontrolOfficialTranscriptProceedings2021}
{Advisory Committee on Reactor Safeguards Digital Instrumentation and Control}.
\newblock Official {{Transcript}} of {{Proceedings Nuclear Regulatory
  Commission}}.
\newblock \url{https://www.nrc.gov/docs/ML2132/ML21320A055.pdf}, October 2021.

\bibitem{andersonWhyInformationSecurity2001}
Ross Anderson.
\newblock Why information security is hard - an economic perspective.
\newblock In {\em Seventeenth {{Annual Computer Security Applications
  Conference}}}, pages 358--365, New Orleans, LA, USA, 2001. IEEE Comput. Soc.
\newblock \url{http://ieeexplore.ieee.org/document/991552/}.

\bibitem{lidi}
ANSSI.
\newblock {lidi}, Feb 2026.
\newblock \url{https://github.com/ANSSI-FR/lidi}.

\bibitem{lidiCLI}
ANSSI.
\newblock {lidi Command line parameters}, Jan 2026.
\newblock \url{https://anssi-fr.github.io/lidi/parameters.html}.

\bibitem{anthropicProjectGlasswingSecuring2026}
{Anthropic}.
\newblock Project {{Glasswing}}: {{Securing}} critical software for the {{AI}}
  era, April 2026.
\newblock \url{https://www.anthropic.com/glasswing}.

\bibitem{PowerEfficiency}
{Apple}.
\newblock {Power Efficiency in OS X}, October 2013.
\newblock
  \url{https://www.apple.com/media/us/osx/2013/docs/OSX_Power_Efficiency_Technology_Overview.pdf}.

\bibitem{DispatchQoS}
{Apple}.
\newblock {DispatchQoS}, May 2026.
\newblock \url{https://developer.apple.com/documentation/dispatch/dispatchqos}.

\bibitem{arnoldStrategiesTransportingData2016}
Ross~D. Arnold.
\newblock Strategies for {{Transporting Data Between Classified}} and
  {{Unclassified Networks}}.
\newblock Technical report, Defense Technical Information Center, Fort Belvoir,
  VA, March 2016.
\newblock \url{https://apps.dtic.mil/sti/citations/AD1005160}.

\bibitem{barryDataDiodesCyber2012}
Courtney Barry.
\newblock Data {{Diodes}} for {{Cyber Security}}.
\newblock {\em NRECA Cooperative Research Network TechSurveillance Magazine},
  March 2012.
\newblock
  \url{https://web.archive.org/web/20231220100501/http://courtneybarry.com/Images/TS_Data_Diodes.pdf}.

\bibitem{bellLookingBackBellLa2005}
David~Elliott Bell.
\newblock Looking {{Back}} at the {{Bell-La Padula Model}}.
\newblock In {\em 21st {{Annual Computer Security Applications Conference}}
  ({{ACSAC}}'05)}, pages 337--351, Tucson, AZ, USA, 2005. IEEE.
\newblock \url{http://ieeexplore.ieee.org/document/1565261/}.

\bibitem{bellSecureComputerSystems1973}
David~Elliott Bell and Leonard~J. LaPadula.
\newblock Secure {{Computer Systems}}: {{Mathematical Foundations}}.
\newblock Technical Report MTR-2547, The MITRE Corporation, March 1973.
\newblock \url{https://apps.dtic.mil/sti/tr/pdf/AD0770768.pdf}.

\bibitem{bergemannCyberSecurityEvent2015}
Brad Bergemann.
\newblock Cyber {{Security Event Notifications}}.
\newblock Technical Report Regulatory Guide 5.83, U.S. Nuclear Regulatory
  Commission, July 2015.

\bibitem{bergmanBattleWorldMost2022}
Ronen Bergman and Mark Mazzetti.
\newblock The {{Battle}} for the {{World}}'s {{Most Powerful Cyberweapon}}.
\newblock {\em The New York Times}, January 2022.
\newblock
  \url{https://www.nytimes.com/2022/01/28/magazine/nso-group-israel-spyware.html}.

\bibitem{berretGuideSecureDrop2016}
Charles Berret.
\newblock Guide to {{SecureDrop}}.
\newblock Technical report, Tow Center for Digital Journalism, 2016.

\bibitem{blanchetBadUSBThreatHidden2018}
St{\'e}phanie Blanchet.
\newblock {{BadUSB}}, the threat hidden in ordinary objects.
\newblock Technical report, Bertin Technologies, June 2018.

\bibitem{hairgap}
CEA.
\newblock {Hairgap}, Apr 2017.
\newblock \url{https://github.com/cea-sec/hairgap}.

\bibitem{cohenDesigningProvablyCorrect1988}
Fred Cohen.
\newblock Designing provably correct information networks with digital diodes.
\newblock {\em Computers \& Security}, 7(3):279--286, June 1988.
\newblock
  \url{https://www.sciencedirect.com/science/article/abs/pii/016740488890034X}.

\bibitem{downsCyberSecurityPrograms2017}
James Downs.
\newblock Cyber {{Security Programs For Nuclear Fuel Cycle Facilities}}.
\newblock Technical Report Draft Regulatory Guide DG-5062, U.S. Nuclear
  Regulatory Commission, January 2017.

\bibitem{fendIndustries}
Fend.
\newblock {Industries}, May 2026.
\newblock \url{https://www.fend.tech/industries}.

\bibitem{fendArchive}
{Fend Incorporated}.
\newblock {Fend XE15 Data Diode}, Mar 2023.
\newblock
  \url{https://web.archive.org/web/20230315012223/https://www.fend.tech/fend-xe15-data-diode}.

\bibitem{fidlerZeroProgressZero2024}
Mailyn Fidler.
\newblock Zero {{Progress}} on {{Zero Days}}: {{How}} the {{Last Ten Years
  Created}} the {{Modern Spyware Market}}.
\newblock {\em Nebraska Law Review}, 103, May 2024.

\bibitem{goodinIOS0daysCellular2023}
Dan Goodin.
\newblock 3 {{iOS}} 0-days, a cellular network compromise, and {{HTTP}} used to
  infect an {{iPhone}}, September 2023.
\newblock
  \url{https://arstechnica.com/security/2023/09/how-the-iphone-of-a-presidential-candidate-in-egypt-got-hacked-for-the-2nd-time/}.

\bibitem{androidThreadPriority}
{Google}.
\newblock {Android Thread MAX\_PRIORITY}, May 2026.
\newblock
  \url{https://developer.android.com/reference/java/lang/Thread\#MAX_PRIORITY}.

\bibitem{sendOnlyEthernet}
Dmitry Grigoryev.
\newblock {Implement send-only (one-way) Ethernet cable}, Dec 2017.
\newblock \url{https://electronics.stackexchange.com/a/279277}.

\bibitem{harpCS2AIKPMGControlSystem2024}
Derek Harp, Bengt {Gregory-Brown}, Walter Risi, and Andrew Ginter.
\newblock The ({{CS}}){{2AI-KPMG Control System Cybersecurity Annual Report}}.
\newblock Technical report, Control System Cyber Security Association
  International, March 2024.

\bibitem{NLDOSDD}
Cyber~Innovation Hub.
\newblock {The Open Source Data Diode}, Nov 2022.
\newblock \url{https://github.com/CyberInnovationHub-NLD/OpenSourceDataDiode}.

\bibitem{industrialcontrolsystemscyberemergencyresponseteamRecommendedPracticeImproving2016}
{Industrial Control Systems Cyber Emergency Response Team}.
\newblock Recommended {{Practice}}: {{Improving Industrial Control System
  Cybersecurity}} with {{Defense-in-Depth Strategies}}, September 2016.

\bibitem{kennethIntegrityConsiderationsSecure1977}
Biba~J. Kenneth.
\newblock Integrity considerations for secure computer systems.
\newblock Technical Report MTR-3153, The MITRE Corporation, Bedford, MA, June
  1977.
\newblock \url{https://apps.dtic.mil/sti/tr/pdf/ADA039324.pdf}.

\bibitem{godiode}
klockcykel.
\newblock {DIY Data Diode}, Sep 2024.
\newblock \url{https://github.com/klockcykel/godiode}.

\bibitem{udpcastCMD}
Alain Knaff.
\newblock {UDPcast commandline options}, Jan 2012.
\newblock \url{http://www.udpcast.linux.lu/cmd.html}.

\bibitem{udpcast}
Alain Knaff.
\newblock {UDPcast}, May 2026.
\newblock \url{http://www.udpcast.linux.lu}.

\bibitem{lagadecDiodeReseauExeFilter2006}
Philippe Lagadec.
\newblock {Diode r\'eseau et ExeFilter : 2 projets pour des interconnexions
  s\'ecuris\'ees}.
\newblock {\em Proceedings of SSTIC06}, 2006.

\bibitem{larkinSecuringPhotovoltaicSystem2020}
Robert~D. Larkin, Torrey~J. Wagner, and Barry~E. Mullins.
\newblock Securing {{Photovoltaic System Deployments}} with {{Data Diodes}}.
\newblock In {\em 2020 47th {{IEEE Photovoltaic Specialists Conference}}
  ({{PVSC}})}, pages 2525--2531, Calgary, AB, Canada, June 2020. IEEE.
\newblock \url{https://ieeexplore.ieee.org/document/9300863/}.

\bibitem{linResearchPacketLoss2013}
Honggang Lin.
\newblock Research on {{Packet Loss Issues}} in {{Unidirectional
  Transmission}}.
\newblock {\em Journal of Computers}, 8(10):2664--2671, October 2013.
\newblock
  \url{https://web.archive.org/web/20240415211830/http://www.jcomputers.us/vol8/jcp0810-29.pdf}.

\bibitem{luBADUSBCRevisitingBadUSB2021}
Hongyi Lu, Yechang Wu, Shuqing Li, You Lin, Chaozu Zhang, and Fengwei Zhang.
\newblock {{BADUSB-C}}: {{Revisiting BadUSB}} with {{Type-C}}.
\newblock {\em 2021 IEEE Security and Privacy Workshops (SPW)}, 2021.

\bibitem{minderRaptorQForwardError2011}
Lorenz Minder, Amin Shokrollahi, Mark Watson, Michael Luby, and Thomas
  Stockhammer.
\newblock {{RaptorQ Forward Error Correction Scheme}} for {{Object Delivery}}.
\newblock Request for {{Comments}} RFC 6330, Internet Engineering Task Force,
  August 2011.
\newblock \url{https://datatracker.ietf.org/doc/rfc6330}.

\bibitem{officeofnuclearregulatoryresearchCyberSecurityPrograms2010}
{Office of Nuclear Regulatory Research}.
\newblock Cyber {{Security Programs For Nuclear Facilities}}.
\newblock Technical Report Regulatory Guide 5.71, U.S. Nuclear Regulatory
  Commission, January 2010.

\bibitem{tfc}
Markus Ottela.
\newblock {Tinfoil Chat}, Apr 2023.
\newblock \url{https://github.com/maqp/tfc}.

\bibitem{owlcyberdefense}
{Owl Cyber Defense}.
\newblock {Data Diode Cybersecurity Products}, May 2026.
\newblock \url{https://owlcyberdefense.com/products/data-diode-products/}.

\bibitem{pietre-cambacedesDeconstructionIndustrialControl2009}
Ludovic {Pi{\`e}tre-Cambac{\'e}d{\`e}s} and Pascal Sitbon.
\newblock Deconstruction of some industrial control systems cybersecurity
  myths.
\newblock {\em Sixth American Nuclear Society International Topical Meeting on
  Nuclear Plant Instrumentation, Control, and Human-Machine Interface
  Technologies}, 2009.

\bibitem{qualcommRaptorQTechnicalOverview2010}
{QUALCOMM}.
\newblock {{RaptorQ}}™ {{Technical Overview}}.
\newblock Technical report, 2010.
\newblock
  \url{https://www.qualcomm.com/content/dam/qcomm-martech/dm-assets/documents/RaptorQ_Technical_Overview.pdf}.

\bibitem{RaptorQPatents}
{Qualcomm Incorporated}.
\newblock {RaptorQ Patents}, May 2026.
\newblock
  \url{https://patents.google.com/?q=(raptorq)&assignee=Qualcomm+Incorporated&num=100}.

\bibitem{rizzoEffectiveErasureCodes1997}
Luigi Rizzo.
\newblock Effective erasure codes for reliable computer communication
  protocols.
\newblock {\em ACM SIGCOMM Computer Communication Review}, 27(2):24--36, April
  1997.
\newblock \url{https://dl.acm.org/doi/10.1145/263876.263881}.

\bibitem{schneierHowAIChanging2026}
Bruce Schneier and Barath Raghavan.
\newblock How {{AI Is Changing Cybersecurity}}, April 2026.
\newblock \url{https://spectrum.ieee.org/ai-cybersecurity-mythos}.

\bibitem{seeberg}
Sven Seeberg.
\newblock {A Data Diode with 2 Raspberry Pi and OpenBSD}, Sep 2025.
\newblock \url{https://github.com/svenseeberg/data-diode}.

\bibitem{siemens}
Siemens.
\newblock {New Siemens data diode now available: secure monitoring of your
  networks}, Dec 2017.
\newblock
  \url{https://www.mobility.siemens.com/global/en/portfolio/rail/stories/new-siemens-data-diode-now-available-secure-monitoring-of-your-networks.html}.

\bibitem{stevensImplicationOpticalData1999}
Malcolm~W Stevens.
\newblock An {{Implication}} of an {{Optical Data Diode}}.
\newblock Technical Report DSTO-TR-0785, {Information Technology Division
  Electronics and Surveillance Research Laboratory}, 1999.

\bibitem{storyBuildingAffordableData2023}
Peter Story.
\newblock Building an {{Affordable Data Diode}} to {{Protect Journalists}}.
\newblock In {\em Workshop on {{Privacy Engineering}} in {{Practice}} ({{PEP}}
  '23)}, August 2023.
\newblock \url{https://peterstory.me/publications/story_pep_2023.pdf}.

\bibitem{storyDefendingMessagingApps2026}
Peter Story.
\newblock Defending {{Messaging Apps Against Spyware Using Data Diodes}}.
\newblock In {\em Free and {{Open Communications}} on the {{Internet}}},
  number~1, 2026.
\newblock \url{https://www.petsymposium.org/foci/2026/foci-2026-0009.pdf}.

\bibitem{wirehair}
Christopher Taylor.
\newblock {Wirehair}, Dec 2023.
\newblock \url{https://github.com/catid/wirehair}.

\bibitem{errorReceived}
{The Python Software Foundation}.
\newblock {Transports and Protocols}, May 2026.
\newblock
  \url{https://docs.python.org/3/library/asyncio-protocol.html\#asyncio.DatagramProtocol.error_received}.

\bibitem{sendtoMacOS}
user2085689.
\newblock {sendto() dgrams do not block for ENOBUFS on OSX}, May 2013.
\newblock
  \url{https://stackoverflow.com/questions/16555101/sendto-dgrams-do-not-block-for-enobufs-on-osx}.

\bibitem{voutevaFeasibilityDeploymentBad2015}
Stella Vouteva, Ruud Verbij, and Jarno Roos.
\newblock {\em Feasibility and {{Deployment}} of {{Bad USB}}}.
\newblock System and {{Network Engineering Master Research Project}},
  University of Amsterdam, February 2015.

\bibitem{OSDD}
Vrolijk.
\newblock {Get started with Data Diodes}, Jan 2026.
\newblock \url{https://github.com/Vrolijk/OSDD}.

\bibitem{DYODE}
Wavestone.
\newblock {Do Your Own Diode}, Jan 2020.
\newblock \url{https://github.com/wavestone-cdt/dyode}.

\bibitem{autonegotiation}
Wikipedia.
\newblock {autonegotiation}, May 2026.
\newblock \url{https://en.wikipedia.org/wiki/Autonegotiation}.

\bibitem{netcat}
Wikipedia.
\newblock {netcat}, May 2026.
\newblock \url{https://en.wikipedia.org/wiki/Netcat}.

\bibitem{zerodiumArchive}
{Zerodium}.
\newblock Zerodium {{Exploit Acquisition Program}}, December 2024.
\newblock
  \url{https://web.archive.org/web/20241217190417/https://zerodium.com/program.html}.

\end{thebibliography}
